# An improved CF-MAC protocol for VANET

**Ghassan Samara**
Computer Science Department, Faculty of Information Technology, Zarqa University, Jordan

**ABSTRACT**

Vehicular Ad hoc Network (VANET) is one of the emerging research areas in the mobile computing field which is considered as future technology and promising topic in computer science and computer networks. Which provides road safety, updated traffic information, and infotainment. VANET consists of a large number of vehicles moving in high speeds while broadcasting important information like safety and control information which must be sent with high priority. Crowded networks like VANET having many vehicles competing to reserve the channel to send critical information which may lead to high collision scenarios, and therefore, there must be a protocol to send this kind of information with high reliability, low data loss and with no collision. In this research a collision-free protocol will be proposed to manage the channel access among competing vehicles to eliminate the collisions which occur rapidly in VANET especially in dense situations, the proposed protocol hereinafter will be called (I-MAC) protocol expected to enhance the channel performance, achieve load balancing, fairness, and decrease message loss and enhance reliability, The evaluation criteria will examine the channel throughput, message delay, and message loss; the results show that the overall channel performance with regard to collision and packet loss ratio is improved.

*Corresponding Author:*

Ghassan Samara,
Computer Science Department,
Faculty of Information Technology,
Zarqa University,
Zarqa, Jordan.
Email: gsamara@zu.edu.jo

## 1. INTRODUCTION

Vehicular ad hoc networks (VANETs) is an important research field for Intelligent Transportation Systems (ITS). Based on IEEE 802.11p [2], VANET has its special wireless radios called Dedicated Short Range Communication (DSRC) that support high-speed communication, high mobile nodes, and high bandwidth. The US Federal Communication Commission (FCC) has allocated 75MHz of the spectrum at 5.9 GHz dedicated for vehicular networks. VANET channel is divided into seven channels with 10 MHz for each [3]. The Channel 178 is the Control Channel (CCH) for the transmission of emergency and status messages (Beacon). The other channels are Service Channels (SCHs), where more information transfers and some applications. Channel 172 and Channel 184 are utilized for safety applications in Vehicle-to-Vehicle (V2V) communication.

VANET channel is isolated into seven 10 MHz channels [3]. The Channel 178 is the Control Channel (CCH) for the transmission of crisis and status messages (Beacon). Alternative channels are Service Channels (SCHs), where more information exchanges and unique applications. Channel 172 and Channel 184 are used in the vehicle to vehicle (V2V) correspondence for well - being applications.

Data transfer rate can reach 54 Mbps. Many single-channel [4-6] MAC protocols have been developed for VANET, where all vehicles are competing for the reservation of a shared single network communication channel. MAC Wireless Access in Vehicle Environments (WAVE) uses the Carrier Sense Multiple Access with Collision Avoidance (CSMA / CA) system to implement contention - based multiple access. As high collision in dense situations and this leads to the channel to collapse quickly.

The Vehicle Ad - hoc Network (VANET) is an extremely dynamic network consisting of a large number of connected mobile vehicles. VANET is a form of Mobile Ad hoc Networks (MANET), and it has no centralized access point. VANET aims to support many applications for safety, entertainment, and traffic optimization. VANET vehicles usually equipped with communication device like Global Positioning System (GPS) receiver, On-Board Unit (OBU), and a set of stationary units along roads, called Road Side Units (RSUs). Based on OBU and RSU, VANET has two key communications: Vehicle-to-Vehicle (V2V) in which vehicles communicate with each other on the move and Vehicle-to-RSU (V2R) where the moving vehicle communicate with the fixed RSU.

Moving Vehicles on the roads are forming a platoon phenomenon; many vehicles at the same location mean more contention, delay, and possibly high collision. Many applications have been developed in the recent years (e.g., See [7]) given the idea of vehicles exchanging Cooperative Awareness Messages (CAM) or beacons, to improve traffic efficiency and safety or to give infotainment. In particular situations where delay message delivery can be highly necessary. One of the reasons why the IEEE 802.11p standardization is adopted of [8, 9] is that the behavior of the 802.11 family is well known.

Beacons are transmitted with ten messages per second using the CSMA / CA Broadcast method [10]. It is thus important to properly manage MAC layer behavior. The beaconing model design was presented in [11] and enclosed the entire saturation spectrum. IEEE 802.11, was designed for wireless LAN (WLAN), and has two disadvantages in its medium access control (MAC) sublayer mechanism (CSMA) carrier sense multiple access: as it might cause a delay in channel access, the other issue is the channel collision. All vehicles within the channel must reserve for transmission, provided that the channel is busy with a different broadcast, all vehicles must wait until this broadcast ends.

The MAC protocol shall be used to prevent or detect a collision between the nodes in order to reserve the channel, and hence this protocol decides who is going to reserve the channel to transmit. In a carrier sense system, such as CSMA, each sending vehicle first listens to the channel to see whether its idle and no other vehicle is currently transmitting, so, if it is idle, the vehicle transmits directly, in the meantime, another vehicle is likely to transmit, and this results in a channel collision.

Additionally, the vehicle may encounter long channel access delays while listening to a busy channel. These two cases occur mostly in dense situations and in networks like VANET. CSMA is usually utilized by IEEE802.11 and IEEE 802.3 Ethernet. The direct implementation of the standard plus reasonable equipment prices is one of the reasons for wide use for WiFi and Ethernet. Due to this Wi-Fi is sometimes used in networks that were not originally designed for. Although CSMA is not adequate in realtime due to high delays in accessing the channels, ethernet ended up being commonly used in the industry where many real-time systems were needed.

Nevertheless, the MAC problems can be solved through the use of more network devices such as switches and routers; therefore the number of vehicles competing for the channel is reduced, in other words, collisions are decreased. However, there is no simple solution in the wireless domain because all vehicles share the wireless channel.

In addition, when using CSMA in the wireless environment, a vehicle can jam the network; even if this vehicle is not doing a real communication, hence, the wireless environment for the carrier sense system is more likely to have interference as there will be no access for other vehicles as long as the transmission on the channel is detected.

The IEEE 802.11p standard, which is intended for VANET environment [12], utilizes CSMA as its MAC access, despite the knowledge that CSMA causes high - mobile systems issues, A solution for CSMA delays could be the use of a self - organized multiple access time division (STDMA), a decentralized, predictable, MAC protocol with limited channel access delay, make it suitable to real-time VANETs.

The proposed research aims to improve the collision-free protocol CF-MAC which was proposed in [1] aiming to enhance channel performance, achieve load balancing, fairness, and decrease message loss and enhance reliability.

## 2. THE CF-MAC PROTOCOL

The CF - MAC protocol was implemented in [1], and works as follows: Only vehicle to vehicle V2V communication infrastructure will be engaged, no roadside units [13]. Each vehicle creates an updated Neighbor Table (NT) containing all neighboring vehicles ' ID and MAC addresses, fresh information about the surrounding network is provided by the received neighbor's beacons, see table 1.

A dynamic TDMA is used to manage the channel access. With no Carrier Sense Multiple Access/ Collision Avoidance CSMA/CA to detect the collision and the transmission in the channel [13-15] as collision happens at the receiver, not at the sender. Platoon shape phenomenon is used [16] and [17] as adjacent vehicles sharing common channel properties, and same neighbor table which includes close vehicles which contend with the current node to reserve the channel, see Figure 1 for the platoon and Table 1 for the Neighbors' Table.

Table 1. Neighbor table structure

| MAC Address | Longitude | Latitude |
|---|---|---|
| 00-1D-0F-C3-01-D3-1D-0F | 40.689060 | 74.152456 |
| 01-1D-3D-C3-78-F6-1D-3D | 40.344536 | 74.341231 |
| 02-1D-0F-R6-42-F2-A1-33 | 41.752370 | 75.000040 |
| 03-1D-0F-C3-77-F7-11-T8 | 42.601021 | 74.753527 |

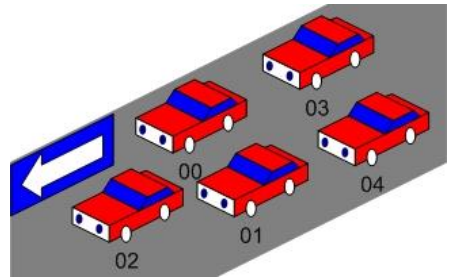

Figure 1. Platoon shape

### 2.1. The CF-MAC protocol steps

Vehicles create Initial Broadcast Table (IBT) before sending a message the IBT is ordered according to the neighboring vehicle MAC address, the MAC address is extracted from local NT, and first slots are given priority for transmission, the IBT initiator will add its MAC address at the first to get the highest priority, see Table 2 which illustrates the IBT structure.

Table 2. Initial broadcast table

| Vehicle MAC | WTS |
|---|---|
| 01-1D-3D-C3-78-F6-1D-3D | |
| 02-1D-0F-R6-42-F2-A1-33 | |
| 00-1D-0F-C3-01-D3-1D-0F | |
| 03-1D-0F-C3-77-F7-11-T8 | |

The created IBT will be broadcasted, so when all neighbor vehicles receive the IBT, they will know their order for transmission, the benefit of IBT that all nearby vehicles will know about their turn and there will be no collision. When a vehicle receives an IBT from a neighbor, it will read the first slot MAC to identify initiator, If this vehicle wants to send, reply to the initiator with a small message (WTS), after receiving WTS reply messages from neighbors, the initiator modifies the IBT by placing 1 in the vehicles' WTS slot for each vehicle made a reply, afterward, broadcasts the IBT to the neighbors, then all the neighbors know who's going to send. For details on how the protocol works, see figures 2 and 3. Where vehicle 01 initiates an IBT and broadcasts, vehicles (00, 02, 03) reply if they want to transmit during this time slot by sending WTS message to vehicle 01, see Figure 3.

The vehicle which has reserved the first slot at IBT and which is currently the initiator has the sending priority, So this vehicle begins to send its message, in the sent message IBT will be piggybacked. When a receiver receives the message, it extracts it together with the IBT. The IBT will show who will send the message next. All vehicles in the nearby network have a knowledge of the IBT transmission priority, so each vehicle will transmit in its turn by waiting for the vehicle that lies before it in the IBT to finish the transmission, see the BT after the ordering in Table 3.

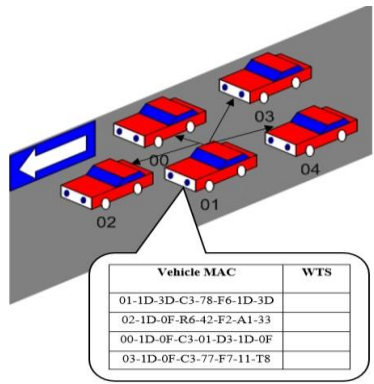
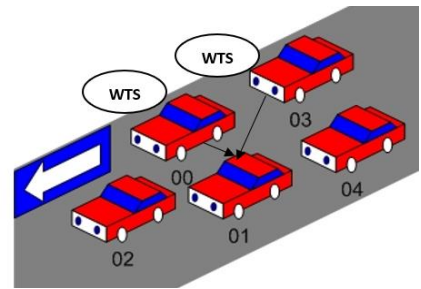

Figure 2. Broadcast table	Figure 3. Want to send

Table 3. Broadcast Table

| Vehicle MAC | WTS |
|---|---|
| 01-1D-3D-C3-78-F6-1D-3D | 1 |
| 02-1D-0F-R6-42-F2-A1-33 | 0 |
| 00-1D-0F-C3-01-D3-1D-0F | 1 |
| 03-1D-0F-C3-77-F7-11-T8 | 1 |

In the last example, vehicle 01 will begin the transmission, and it will include the final BT appended (piggybacked) with the transmitted message, and vehicle 00 will wait for the finish of the transmission, at that moment, vehicle 03 is waiting for vehicle 00 transmission, after the finish of vehicle 03 transmission, the first slot that has 0 in WTS is permitted to transmit its BT, and in this way, vehicle 02 will begin transmitting its BT which contains fresh information about the vehicles neighborhood.

In case there is no vehicle has 0 in its WTS, any vehicle inside the BT has a message to send it must compete with the other BT vehicles that want to send, The vehicle will wait a random amount of time with Short Interframe Spacing (SIFS) and Distributed Interframe Spacing (DIFS) during contention time to avoid a collision.

For new vehicles joining the platoon the message should be received, and the neighbors should be contacted once they are asked to do so, It can not start sending as soon as it enters the group, so it will receive the transmission from neighbors, as soon as it receives the BT and want to transmit, it will include 1 in its WTS slot and waits for its turn to begin transmitting.

## 3.  THE I-MAC PROTOCOL

Although the CF - MAC has good results, improvements are still possible, the proposed I-MAC will add the following: The initiator must use Carrier Sense Multiple Access (CSMA) before sending the IBT, Since there is a possibility of several vehicles simultaneously sending IBT and thus collisions occur, so, the vehicle will listen to the channel before sending the IBT, Using CSMA is reasonable in the current situation as the vehicles are close as a consequence of the platoons, so, if the channel is busy, vehicle will wait until the channel becomes idle before transmission.

In addition, many collisions occur if the vehicle transmits immediately when sensing an idle channel, so fairness method must be applied, and in here, DIFS will be used, DIFS means vehicle has to wait a random amount of time  when sensing an idle channel before sending, and this is longer than SIFS, As the use of SIFS in a crowded network like VANET, many vehicles wait very short and more than one vehicle still has a chance to stop waiting and start transmission simultaneously. See Figure 4 for the I-MAC protocol flowchart

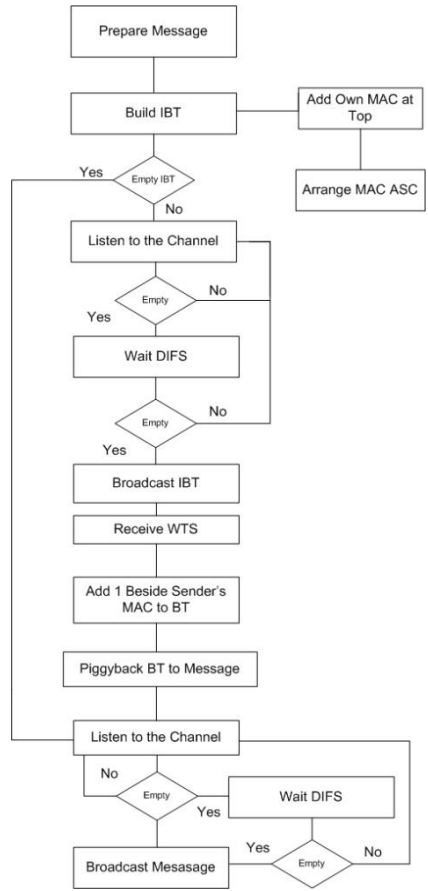

Figure 4. I-MAC protocol flowchart

## 4. SIMULATION AND RESULTS

Matlab R2018b was used in this paper to test the correctness of the proposed I - MAC protocol, the protocol tested and compared with the CF-MAC [1] and DTMAC [18], in the I - MAC improved protocol, where the precise environment and all the simulation parameters used in CF - MAC and DTMAC were adopted.

### 4.1. Simulation parameters

Highway scenarios with two lanes of size 2000m × 20m, in both directions, the speed of the moving vehicle is constant; the simulation parameters are summarized in Table 4. Nakagami propagation model was utilized to distribute the network nodes [17].

Table 4. Simulation parameters

| Street length | 2 km |
|---|---|
| Lanes | Two lanes |
| Speed | 120 km/h |
| Transmission range | 300 m |
| Number of vehicles | 100 |
| Network interface | Phy/WirelessPhyExt |
| MAC interface | Mac/802 11Ext |
| Interface queue | Queue/DSRC |
| Propagation model | Propagation/Nakagami |
| Number of TDMA slots/frames | 10 |
| Time slot | 2.5ms |
| Antenna type | Antenna/omniantenna |
| Minimum beaconing interval | 100ms |
| Maximum beaconing interval | 500ms |

Area occupancy (AO) parameter was used in [18], which is equal to $\frac{N \times R}{L \times T}$ in a highway scenario, where N is the total number of active vehicles, R is the communication range, L is the length of the highway, T is the number of slots reserved for each area, this will indicate the dense situations, as these protocols were tested in high traffic scenarios.

Figure 5 shows the merging collision rates of I-MAC, CF-MAC, and DTMAC. As shown in this Figure, I-MAC achieves less collision than CF-MAC and DTMAC starting from AO > 0.9 where IF-MAC works better as it eliminated the chance for the collision to some degree, furthermore, I-MAC achieves better and consistent performance in higher traffic scenarios than CF-MAC, which makes it more reliable.

For instance, at AO = 1, the IF-MAC protocol achieves in merging collision rate of 3%, CF-MAC scores 9% while which shows a rate of 22% (which means that CF-MAC is higher in collision by 6%, and DTMAC is higher by 19%), it is also worth noting that when OA = 1 means that the area is fully crowded. These results can be explained by the fact that CF-MAC and DTMAC have achieved a higher rate of merging collision compared to I-MAC.

Figure 6 shows the access collision rates of I-MAC, CF-MAC, and DTMAC. As shown in this Figure, IF-MAC achieves a smaller rate of access collisions than CF-MAC and DTMAC starting from AO ($\geq$ 0.8). For instance, at AO = 1, the IF-MAC protocol achieves an access collision rate of 5%, CF-MAC scores 9% while which shows a rate of 71% (which means that CF-MAC is higher in collision by 4%, and DTMAC is higher by 66%), which means when the channel is fully crowded DTMAC can't tolerate the channel access and many vehicles transmit at the same time.

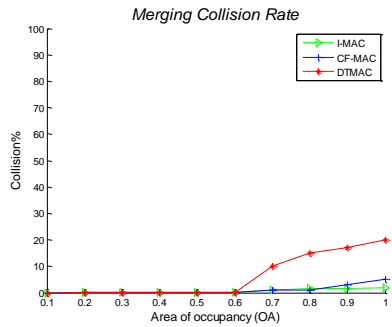

Figure 5. The rate of merging collision

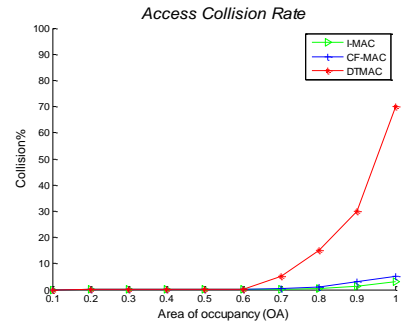

Figure 6. The access collision rate

The packet loss rates of the I-MAC, CF-MAC, and DTMAC protocols are shown in Figure 7. For AO $\leq$ 0.8, the I-MAC outperforms the other two protocols, this means that no packet loss for the I-MAC if the network is dense below the AO > %0.9, if the network is fully crowded AO = 100% then just 0.5% packet is lost, , while the CF-MAC starting to lose packets at AO > 0.7, and when the network is fully crowded i.e. AO = 100% CF-MAC scores 1.7% of packet loss which is much higher than the I-MAC, the DTMAC starting to lose packets at AO > 0.6, and when the network is fully crowded, i.e., AO = 100% DTMAC scores 3.8% of packet loss which is too much higher than the I-MAC. It can be seen that the proposed protocol has the lowest packet loss rate, especially for a high AO, due to its capability to deal with the merging collision problem.

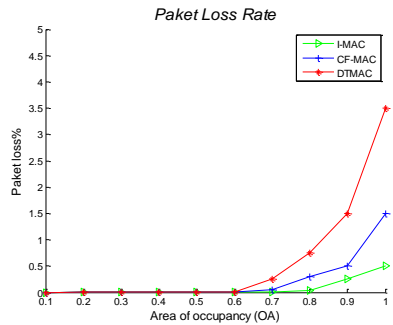

Figure 7. The rate of access collision

## 5. CONCLUSION

This paper proposes a new I - MAC protocol to improve the CF - MAC, this protocol is aimed at controlling channel access (MAC), which in turn improves channel performance and reliability to achieve a collision-free network, The results show that the I - MAC results in terms of collision and the packet loss ratio are performing better compared to CF - MAC and DTMAC.

## ACKNOWLEDGEMENTS

This research is funded by the Deanship of Research and Graduate Studies in Zarqa University, Jordan.